\documentclass[reprint,amsmath,amssymb,aps,prc,showpacs,longbibliography,lengthcheck]{revtex4-1}

\usepackage{epsfig}

\usepackage{amsmath}
\usepackage{graphicx}
\usepackage{amssymb}
\usepackage{bm}
\long\def\symbolfootnote[#1]#2{\begingroup%
\def\thefootnote{\fnsymbol{footnote}}\footnote[#1]{#2}\endgroup}
\begin{document}

\title{Investigation on the $^{48}$Ca+$^{249-252}$Cf reactions
 synthesizing isotopes of the 118 superheavy element}
\author{G. Mandaglio$^{1,2,3}$\symbolfootnote[1]{E-mail: gmandaglio@unime.it},
 G. Giardina$^{2,3}$\symbolfootnote[2]{E-mail: ggiardina@unime.it}, A. K. Nasirov$^{4,5}$,
and A. Sobiczewski$^{6}$}
\affiliation{$^1$Centro Siciliano di Fisica Nucleare e Struttura della Materia,
 95125 Catania, Italy\\
$^2$ Dipartimento di Fisica dell'Universit\`a di Messina, 98166 Messina, Italy\\
$^3$ Istituto Nazionale di Fisica Nucleare, Sezione di Catania, 95123 Catania,  Italy\\
$^4$Joint Institute for Nuclear Research, 141980 Dubna, Russia\\
$^5$Institute of Nuclear Physics, 100214, Tashkent, Uzbekistan\\
$^6$National Centre for Nuclear Research, Ho\.za 69, 00-681 Warsaw, Poland
}
\pacs{25.70.Jj, 25.70.Gh, 25.85.-w}

\begin{abstract}
 The study of the $^{48}$Ca+$^{249,250,251,252}$Cf reactions in a wide energy interval
around the external barrier has been achieved with the aim of investigating the
dynamical effects of the entrance channel via the $^{48}$Ca induced reactions on
the $^{249-252}$Cf targets and to analyze the influence of odd and even neutron
composition in target on the capture, quasifission and fusion cross sections.
Moreover, we also present the results of the individual
evaporation residue excitation functions obtained from the de-excitation cascade of
the various even-odd and even-even $^{297-300}$118 superheavy compound nuclei reached
in the studied reactions, and we compare our results of the  $^{294}$118 evaporation
residue yields obtained in the synthesis process  of the $^{48}$Ca+$^{249,250}$Cf
reactions with the experimental data obtained in the $^{48}$Ca+$^{249}$Cf experiment
carried out at the  Flerov Laboratory of Nuclear Reactions of  Dubna.
\end{abstract}

\date{Today}
\maketitle

\section{Introduction}
\label{intro}

In the last decade experiments were performed by using the $^{48}$Ca
beam against the $^{243}$Am, $^{245,248}$Cm, $^{249}$Bk and
$^{249}$Cf actinide targets (see Refs.
\cite{Ogan05,Ogan04,Ogan10}) in order to synthesize
Z=115,116,117, and 118  superheavy elements, respectively, and to
explore their characteristics. The possibility of
obtaining the heaviest superheavy elements $^{302}$119 and $^{305}$120 by
using the $^{48}$Ca beam in the $^{48}$Ca+$^{254}$Es and
$^{48}$Ca+$^{257}$Fm reactions, respectively is restricted by difficulties
in obtaining a thick enough of $^{254}$Es and $^{257}$Fm actinide targets
because the other Es and Fm isotopes
are radioactive with shorter lifetimes.
Therefore, in order to reach heavier superheavy elements (SHE),
the beams heavier than $^{48}$Ca (as for example $^{50}$Ti,
$^{54}$Cr, $^{58}$Fe, $^{64}$Ni, and other
 heavier projectiles) against the above-mentioned actinide targets
should be used. But, unfortunately, the evaporation residue (ER) cross sections
decreases strongly by decreasing the charge (mass) asymmetry of reactants in
the entrance channel.
 This is connected with the strong hindrance to formation of compound nucleus
  due to  the dominant role of the quasifission process
  which competes with complete fusion.  Quasifission is the decay of the
  formed dinuclear system (DNS) into two fissionlike fragments after
  the charge and/or mass exchange between its components without reaching
  the compound nucleus stage. The capture events that survive quasifission populate the complete fusion formation from which the deformed mononucleus may reach the statistically equilibrated shape of compound nucleus (CN).  Another hindrance to formation of compound nucleus
  appears in collisions with large impact parameter-orbital angular momentum.
  Although DNS can survive against quasifission at large values of angular momentum $L=\ell\hbar$ and
  it should be transformed into complete fusion, the mononucleus still not statically equilibrated can split into two fragments (fast fission process) if the fission barrier of this nuclear system
  disappears for high values of $\ell$ ($B_{\rm f}(\ell>\ell_{\rm f})=0$). Therefore, the fast fission process is present in reactions only at high angular momentum values ($\ell>\ell_{\rm f}$, where  $\ell_{\rm f}$ is a characteristic value for each nucleus), while the quasifission process takes place at all  $\ell$ values contributing to the capture reaction.

  The first experiment which were performed at Flerov Laboratory of
  Nuclear Reaction of Joint Institute for Nuclear Reaction ($^{58}$Fe+$^{244}$Pu
 \cite{Ogan09})
  and at GSI of Darmstadt ($^{64}$Ni+$^{238}$U and $^{54}$Cr+$^{248}$Cm \cite{Hof11},
  and $^{50}$Ti+$^{249}$Cf  \cite{Duel11}) to explore the synthesis of the
  $Z$=120 superheavy element did not identify any event of synthesis of the expected
  superheavy element.
  In our previous papers (see Refs. \cite{NasirovPRC84,NasirovPRC79}), we presented
  results of calculation on the above-mentioned reactions which could lead to the
  $Z$=120 superheavy element, but we found  values of the evaporation residue
 cross sections lower than 0.1 pb. Predictions of other authors are approximately
  near this value \cite{Zagreb117,LiuBao120,AdamAnt,KSiwek,adam86}.
 Therefore, it is necessary to improve the experimental conditions in order to be able to reach measurements of cross sections of the order of fb.
 The dominant role of the quasifission process in reactions
  with massive nuclei is  connected by the increase of the intrinsic fusion barrier
  depending on the shell structure of interacting nuclei and
  rotational energy of DNS which is formed at the given beam energy
  and orbital angular momentum in the entrance channel.
  Moreover, due to the fast fission process taking place at high orbital angular momentum
 values of the complete fusion system and the nearly  fusion-fission process of
 excited and rotating compound nucleus, the possibility to synthesize in future
superheavy elements with $Z>120$ by very massive nuclei reactions appears a
very difficult task. We presented and discussed our results in Refs. \cite{ijmpe09,jpcs2011}
about such perspectives.

The aim of this paper is to study four reactions induced by $^{48}$Ca on the $^{249-252}$Cf
 targets in order to analyze the effect of mass number and structure properties of nuclei in
 the entrance channel on the capture, quasifission, and complete fusion processes.
For that we compare the capture, quasifission, and fusion cross sections for these
considered reactions in order to analyze the influence of the odd or even neutrons present
in target on the above-mentioned cross sections.
The study and comparison of such cross sections allows us to reveal the sensitivity  of
the model and results on the dynamical effects of the entrance channel, while the determination
and analysis of the evaporation residue cross sections for the four reactions reveal the
influence of the different structure of the formed $^{297-300}$118 superheavy compound nuclei
in the $^{48}$Ca+$^{249-252}$Cf reactions with different neutron rich targets.
Such reactions can be expected to occur in  one, exceptional experiment, which could be performed to be performed in a near future. Its specific character should consist in the use of a target which is a mixture of the isotopes $^{249-252}$Cf. The amount of $^{252}$Cf (which has a relatively short half-life) in the target is expected to be very small, but the percentage of the other are thought to be large and comparable with each other.

In Sect. \ref{sect2} we present and analyze the results of the capture, quasifission,
and  fusion cross sections. In Sect. \ref{sect3} we analyze excitation functions of the
evaporation residue (ER) after few neutrons emission only from the formed compound nucleus (CN).
 Moreover, we compare our results for the $^{294}$118 residue nucleus which was
synthesized in the $^{48}$Ca+$^{249,250}$Cf reactions after evaporation of 3 and 4 neutrons from
the $^{297}$118 and $^{298}$118 compound nuclei, respectively, with the experimental
data concerning identification of $^{294}$118 superheavy nucleus observed
in the $^{48}$Ca+$^{249}$Cf experiment reported in Ref. \cite{Ogan06}.
The conclusions are presented in Sect. \ref{sect4}.

\section{Results of the capture, quasifission,  and  fusion cross sections}
\label{sect2}

The evaporation residue formation, fusion-fission, quasifission and fast fission events
take place if capture of projectile-nucleus by target-nucleus occurs
after full momentum transfer due to the friction in relative motion and presence
of potential well in nucleus-nucleus interaction.
\begin{figure}[h]
\vspace*{2.5cm}
\begin{center}
\resizebox{0.52\textwidth}{!}{\includegraphics{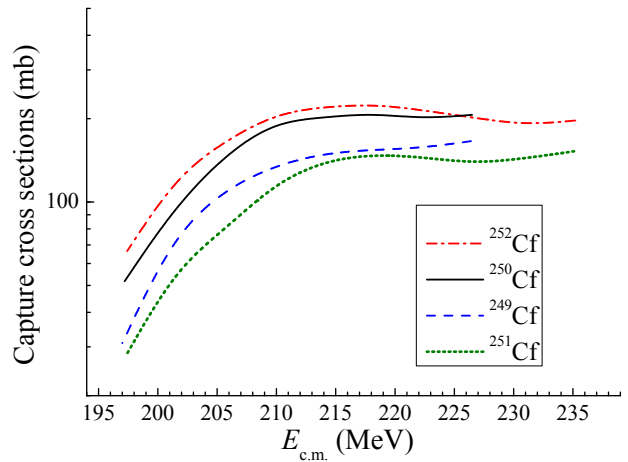}}
\vspace*{-2.55 cm} \caption{\label{fig1} (Color online) Comparison of the capture cross sections for the
$^{48}$Ca+$^{249-252}$Cf reactions.}
\end{center}
\end{figure}

 In this work the capture probability is determined by solving the equations of radial motion,
surface vibration of $^{48}$Ca, and  orbital angular momentum as in
Refs. \cite{GiarSHE,NasirovPRC84,NasirovPRC79,NasirovRauis}:
 \begin{eqnarray}
  F(R)  & = & \mu(R,\alpha_T)\ddot R +
 \gamma_{R}(R,\alpha_T)\dot R, \label{EqR}  \\
   F(R,\alpha_T)&=&  -\frac {\partial V(R,\beta_i,\alpha_T)}{\partial R}-\dot R \hspace{-0.3mm}^2 \frac {\partial \mu(R)}{\partial R}, \label{maineq2}\\
   F_{\beta_i}(R)  & = & D_{\beta_i}\ddot \beta_i(t)+
 \gamma_{\beta}(R)\dot \beta_i(\alpha_T,t)+C_{\beta_i}^2 \beta_i\, ,\label{EqTeta}\\
   F_{\beta}(R) & = &
 -\frac {\partial V(R,\beta_i,\alpha_T)}{\partial \beta_i}, \label{FTeta}\\
 \frac{dL}{dt} & = & \gamma_{\theta}(R,\alpha_T)R \times\nonumber\\ & \times & \left(\dot{\theta}
 R -\dot{\theta_1} R_{1eff} -\dot{\theta_2} R_{2eff}\right), \label{EqL} \\
 L_0 & = & J_R(R,\alpha_T) \dot{\theta}+J_1 \dot{\theta_P}+J_2 \dot{\theta_2},\\
 E_{\rm rot} & = & \mu(R,\alpha_T){\dot R}^2/2+\frac{J_R(R,\alpha_T) \dot{\theta_{}}{}^2}2+ \nonumber \\ & + & \frac{J_1
 \dot{\theta_1}^2}2+\frac{J_2 \dot{\theta_2}^2}2\,,
 \end{eqnarray}
 where $R\equiv R(t)$ is the relative motion coordinate; $\dot R(t)$
 is the corresponding velocity; $\alpha_T$ is the orientation
 angle between beam direction and axial symmetry axis of the target (Cf isotope);
$L_0$ ($L_0=\ell_0 \hbar$) and $E_{\rm rot}$ are defined by
 initial conditions; $J_R$ and $\dot\theta$, $J_1$ and
 $\dot\theta_1$, $J_2$ and $\dot\theta_2$ are  moment of inertia
 and  angular velocities of the DNS and its fragments, respectively;
$\gamma_R$ and $\gamma_{\theta}$ are the friction coefficients for the relative
 motion along $R$ and the tangential motion when two nuclei roll on
 each other's surfaces, respectively; $C_{\beta_i}$  and $\gamma_{\beta_i}$,
 and $D_{\beta_i}$ are stiffness, damping and mass coefficients for the surface
 vibrations of $^{48}$Ca, respectively;
 $V(R,\alpha_T)$ is the  nucleus-nucleus
 potential calculated by the double folding procedure \cite{GiarSHE,FazioJPSJ72}.
 Friction coefficients $\gamma_R$ and $\gamma_{\theta}$ depend on the
 shell  structure of interacting nuclei (see Ref. \cite{AdamianPRC56} and
 Appendix  A of Ref. \cite{NuclPhys05}) and orientation angles of their axial symmetry
  axes (if they are deformed). Our calculations showed that ratio between them
  does not change so much during capture trajectory of collision:
$\gamma_{\theta}/\gamma_R=3\cdot 10^{-3}\div 4\cdot 10^{-3}$.
The values of surface vibration coefficients for the quadrupole multipolarity are
given in Ref. \cite{Raman}. The values of damping coefficient $\gamma_{\beta_i}$
for the surface vibration are calculated by the expression presented in Ref.
\cite{NasirovRauis}. The deformation parameters in the ground state of $^{249-252}$Cf
were obtained from Ref. \cite{MolNix}.

\begin{figure}[h]
\vspace*{2.0cm}
\begin{center}
\resizebox{0.52\textwidth}{!}{\includegraphics{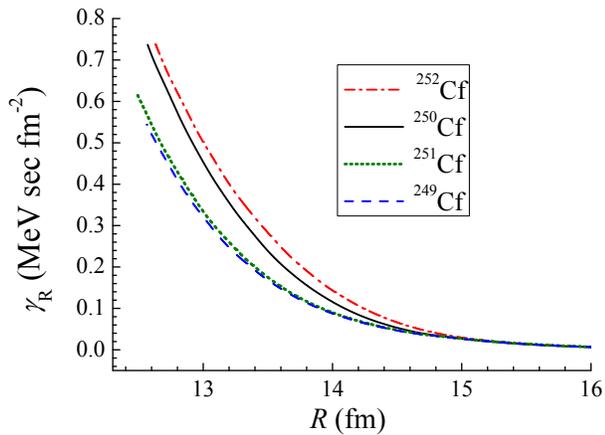}}
\vspace*{-2.55 cm} \caption{\label{fig2} (Color online) Comparison of the radial friction coefficient
 $\gamma_R$ for the $^{48}$Ca+$^{249-252}$Cf reactions. The presented results were
obtained for the orientation angle $\alpha=30^{\circ}$ of the axial symmetry axis of Cf isotopes.}
\end{center}

\end{figure}
\begin{figure}[h]
\vspace*{1.0cm}
\begin{center}
\resizebox{0.52\textwidth}{!}{\includegraphics{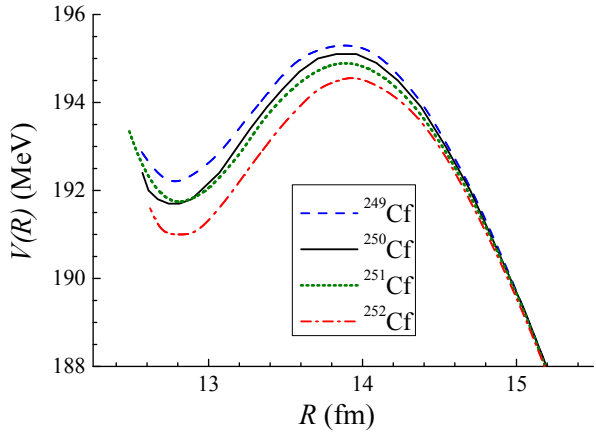}}
\vspace*{-2.55 cm} \caption{\label{fig3} (Color online) Comparison of the nucleus-nucleus potential
$V(R)$ for the $^{48}$Ca+$^{249-252}$Cf reactions. The presented results were obtained
for the orientation angle $\alpha=30^{\circ}$ of the axial symmetry axis of Cf isotopes.}
\end{center}
\end{figure}

 The results of calculation are partial capture cross sections corresponding to
 the angular momentum distribution of the DNS formed after full momentum transfer.
 The probability of capture event is sensitive to the nucleus-nucleus potential and
 friction coefficient which depend on the single-particle states of protons and neutrons in the
 interacting nuclei \cite{AdamianPRC56}.  The last quantities are
 sensitive to the mass number of the given isotope.
 In Fig. \ref{fig1} we compare  the capture cross sections calculated for the
 $^{48}$Ca+$^{249-252}$Cf reactions which can lead to formation of compound nuclei
 being appeared as isotopes of the 118 superheavy element with mass numbers
 $A=297, 298, 299,$ and  $300$.

\begin{figure}[h]
\vspace*{2.5cm}
\begin{center}
\resizebox{0.52\textwidth}{!}{\includegraphics{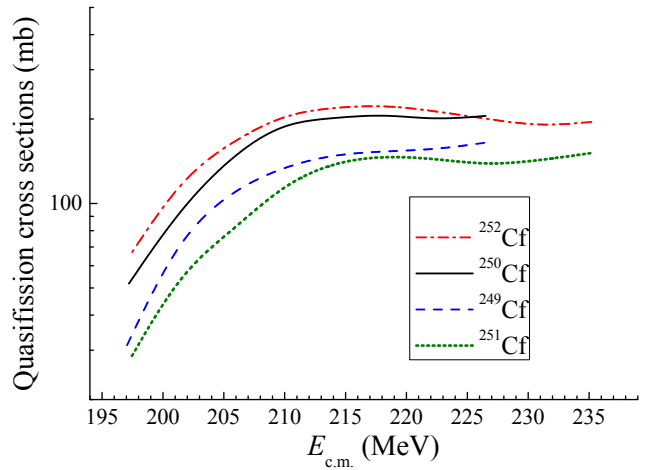}}
\vspace*{-2.30 cm} \caption{\label{fig4} (Color online) Comparison of the quasifission cross sections
for the $^{48}$Ca+$^{249-252}$Cf reactions.}
\end{center}
\end{figure}

\begin{figure}[h]
\vspace*{2.5cm}
\begin{center}
\resizebox{0.52\textwidth}{!}{\includegraphics{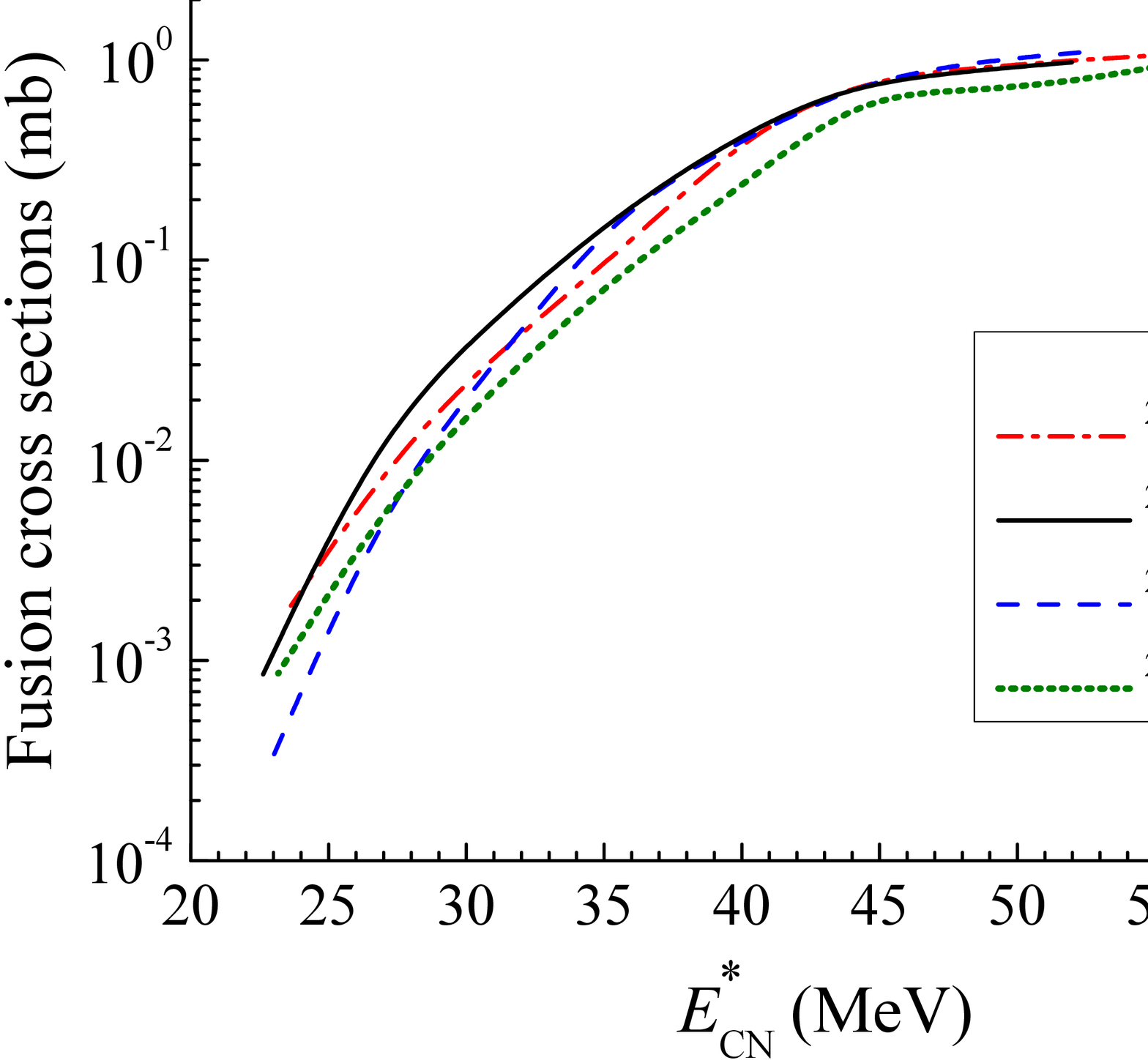}}
\vspace*{-2.60 cm} \caption{\label{fig5} (Color online) Comparison of the fusion cross sections for
the $^{48}$Ca+$^{249-252}$Cf reactions.}
\end{center}
\end{figure}

\begin{figure}[h]
\vspace*{2.0cm}
\begin{center}
\resizebox{0.52\textwidth}{!}{\includegraphics{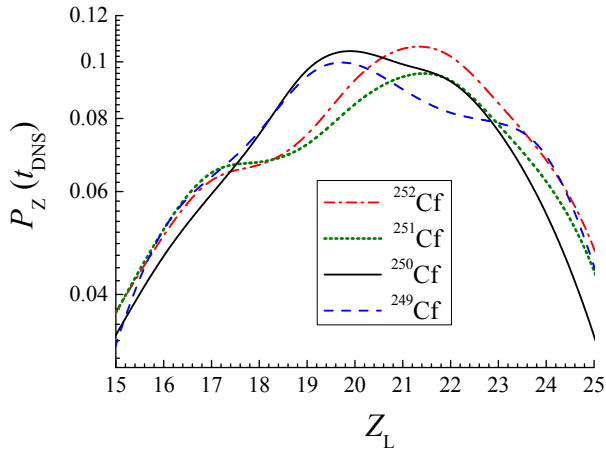}}
\vspace*{-2.55 cm} \caption{\label{fig6} (Color online) Comparison of the behaviour of the charge
distributions between constituents of dinuclear system formed in the $^{48}$Ca+$^{249-252}$Cf
reactions. The results are obtained for the evolution of dinuclear system during
$t_{\rm DNS}=3 \cdot 10^{-21}$ s.}
\end{center}
\end{figure}

\begin{figure}[h]
\vspace*{2.0cm}
\begin{center}
\resizebox{0.52\textwidth}{!}{\includegraphics{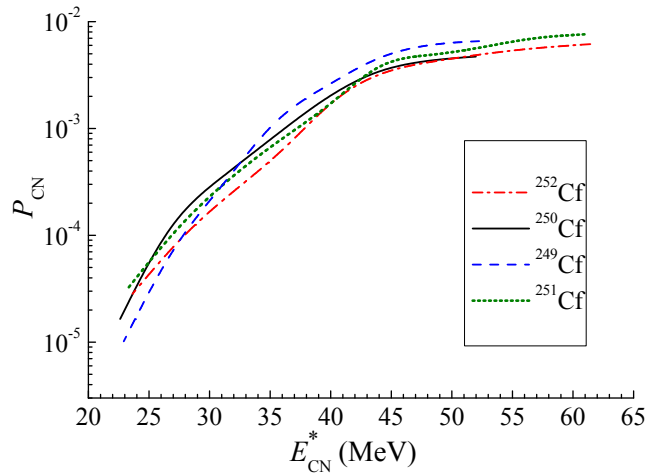}}
\vspace*{-2.60 cm} \caption{\label{fig7} (Color online) Comparison of the $P_{\rm CN}$ fusion probability
for the $^{48}$Ca+$^{249-252}$Cf reactions.}
\end{center}
\end{figure}

 The study of the effect of mass number $A$ on the capture process is reduced to
 analysize the dependence of the friction coefficient and nucleus-nucleus potential
 on $A$.  In Fig. \ref{fig2} we compare the friction coefficients which were calculated
 for the $^{48}$Ca+$^{249-252}$Cf reactions  and for the orientation angle $\alpha_T=30^{\circ}$ of
 the axial symmetry axis of Cf isotopes.
 As one can see the friction coefficients which were obtained for the
 $^{48}$Ca+$^{250,252}$Cf reactions with even isotopes of Cf are appreciably higher than
 the ones calculated for the $^{48}$Ca+$^{249,251}$Cf reactions with its odd isotopes.
 In addition, Fig. \ref{fig3} shows the comparison of the depth of the potential wells
 which are responsible for the capture of nuclei; from this figure appears that in
 reactions with even isotopes of Cf the potential wells are deeper than the ones obtained
 for the reactions with odd isotopes of Cf. The shell effects are included into a
 term $\delta V(R)$ of the nucleus-nucleus potential
\begin{equation}
V(R)=V_{\rm Coul}(R)+V_{\rm nucl}(R)+V_{\rm rot}(R)+\delta V(R),
\end{equation}
 where $V_{\rm Coul}(R)$ is the Coulomb interaction potential which is calculated
 by Wong's formula \cite{Wong}; $V_{\rm nucl}$ is nuclear part which is found by double
 folding procedure with the effective nucleon-nucleon forces \cite{Migdal};
 $V_{\rm rot}$ is DNS rotational energy.
 The nature of $\delta V(R)$ term and friction coefficient $\gamma_R$ are connected
 with nucleon exchange between nuclei \cite{NuclPhys05}:
\begin{equation}
\delta V(R(t)=\Sigma_{i,k}\left|\frac{\partial V_{ik} (R)}{\partial R(t)}\right|^2
B^{(0)}_{ikj}(R(t)),
\end{equation}
\begin{equation}
{\large \gamma}_R=\Sigma_{i,k}\left|\frac{\partial V_{ik} (R)}{\partial R(t)}\right|^2 B^{(1)}_{ikj}(R(t)),
\end{equation}
where
\begin{eqnarray}
 B^{(n)}_{i,k}(R(t))& = & \frac{2}{\hbar}\int_{t_0}^t dt (t-t_0)^n \exp[(t-t_0)/\tau_{ik}]\times \nonumber\\ & \times & \sin [\omega_{ik}(R(t))(t-t_0)](n_i-n_k),
\end{eqnarray}
\begin{equation}
{\rm with}\,\, \,\,\hbar \omega_{ik} = \epsilon_i + \Lambda_{ii}-\epsilon_k-\Lambda_{kk},
\end{equation}
containing the dependence on the single-particle occupation numbers $n_i$ and
energies $\epsilon_k$ of nucleons in the interacting nuclei
($i$ and $k$ states belong to projectile and target nuclei, respectively);
$\Lambda_{ii}$ and $\Lambda_{kk}$ are diagonal elements of the matrix elements
$V_{ik}$ of the DNS meanfield for nucleons;
$n_i$ is the diagonal matrix element of the density matrix which is calculated
according to the model presented elsewhere \cite{AdamianPRC56,AdamPPN25};
\begin{equation}
\tau_{ik} = \frac{\tau_i \cdot \tau_k}{\tau_i + \tau_k},
\end{equation}
$\tau_{i}$ ($\tau_{k}$) is the lifetime of the quasiparticle excitations in the single-particle
state $i$ ($k$) of each reacting  nucleus. It determines the damping of single-particle exitation.
$\tau_{i}$ ($\tau_{k}$) is calculated using the results of the quantum liquid theory
\cite{Pines} and the effective nucleon-nucleon forces from \cite{Migdal}
as in Ref. \cite{AdamianPRC56}.

From this analysis we can conclude that the difference between capture cross sections presented
in Fig. \ref{fig1} is caused by  difference in occupation single-particle states in
 different isotopes of Cf. This difference  quantitatively appears in the friction
coefficient of the radial motion and nucleus-nucleus interaction.

 In calculation of the fusion cross section we use the potential energy surface of
dinuclear system which includes  the binding energies of the interacting nuclei
($B_1$ and $B_2$)  and compound nucleus ($B_{\rm CN}$):
 $U_{\rm dr}=B_1+B_2-B_{\rm CN}+V_{\rm nuc-nuc}$
 where  $V_{\rm nuc-nuc}$ is the nucleus-nucleus interacting potential.
 Therefore, the landscape of the potential energy surface determines competition
between complete fusion and quasifission during evolution of DNS
\citep{NuclPhys05,GiarSHE,FazioJPSJ72}. As discussed in these papers,
 during evolution to compound nucleus, the dinuclear system must overcome the intrinsic
fusion barrier $B^*_{\rm fus}$ which   is determined by its mass and charge asymmetry,
as well as shell structure and orientation angles of symmetry axes of its constituents.
Therefore, the fusion probability of the dinuclear system into compound nucleus
$P_{\rm CN}$ at the given excitation energy $E_{\rm DNS}^*$ is calculated as a
sum of fusion probabilities from different  configuration with different charge asymmetries:
 \begin{eqnarray}
P_{\rm CN}(E_{\rm DNS}^*,\ell, \alpha_T) & = & \sum_{Z=Z_{\rm sym}}^{Z=Z_{\rm max}}
Y_Z(E_{\rm DNS}^{*(Z)},\ell,\alpha_T)\times \nonumber\\ & \times & P^{(Z)}_{\rm CN}(E_{\rm DNS}{*(Z)},\ell,\alpha_T)
 \end{eqnarray}
where $E_{\rm DNS}^* = E_{\rm c.m.} − V(Z,R_m,\ell,\alpha_T)+\Delta Q_{\rm gg}(Z)$
is the excitation energy of DNS for a given value of its charge-asymmetry configuration
$(Z, Z_{\rm tot}- Z)$ and $Z_{\rm tot} = Z_1 + Z_2$; $E_{\rm c.m.}$ is the collision
energy in the center-of-mass system; $V(Z, R_m ,\ell; \alpha_T )$ and $R_m$ are the minimum
value of the nucleus-nucleus potential well and its position on the relative distance
between centers of nuclei; $\Delta Q_{\rm gg} (Z)$ is the change of
$Q_{\rm gg} (Z)-$value by changing the DNS charge asymmetry from the initial
value $Z=Z_1$; $P_Z(E_{\rm DNS}^{*(Z)})$ and $Y_Z(E_{\rm DNS}^{*(Z)})$ are the probabilities
of population of the configuration $(Z, Z_{\rm tot} - Z)$ at $E_{\rm DNS}^{*(Z)}$ and
decay from this configuration, respectively. $Z_{\rm sym} = (Z_1 + Z_2 )/2$ and $Z_{\rm max}$
corresponds to the point where the driving potential reaches its maximum value
$(B_{\rm fus} (Z_{\rm max}) = 0)$ (see Refs. \citep{NuclPhys05,GiarSHE,FazioJPSJ72}).

The theoretical results presented in Fig. \ref{fig4} show that the behaviours of the quasifission
excitation functions for the reactions  under discussion are similar to the corresponding
capture excitation functions shown in Fig. \ref{fig1}, while the  difference fusion excitation
functions (the formalism is shortly presented in Appendix A and references therein)
calculated for these reactions is not so much as for capture excitation function:
 the values of the fusion excitation functions are closer (see Fig. \ref{fig5}).
The reason for which the advance in capture of the  $^{48}$Ca+$^{250,252}$Cf reactions
with even isotopes of Cf   in comparison with the $^{48}$Ca+$^{249,251}$Cf reactions
has been lost is explained by the opposite behaviour of the charge distribution
between constituents of DNS which is formed in the initial stage of the reactions.
As one can see  in Fig. \ref{fig6}  the maximum of the charge distribution in
DNS formed in the reactions with isotopes of Cf with larger ($A$=251 and 252) mass
numbers moves to the charge symmetric direction  while the charge distribution of
ones formed in the reactions with isotopes of Cf with smaller  mass
numbers ($A$=249 and 250) moves to charge asymmetric direction.
 It is known from theoretical models based on the DNS concept \cite{AdamNPA98}
and our calculations that hindrance to
complete fusion increases for the more charge symmetric configurations because in
this case intrinsic fusion barrier increases and quasifission barrier decreases
making system less stable against decay into two fissionlike fragments
(we call them quasifission fragments) \cite{NasirovPRC79}. Therefore,
although the capture excitation function of the $^{48}$Ca+$^{249}$Cf reaction
was lower but its fusion excitation function grows more fastly than the ones of the
other reactions by increasing the beam energy and becomes even higher (Fig. \ref{fig5}).

 The fusion probability $P_{\rm CN}$ (see Fig. \ref{fig7}) determines fusion cross
section at the given capture cross section and, therefore, its behaviour is similar
of the behaviour of the fusion excitation function presented in Fig. \ref{fig5}.
The close values of the calculated fusion cross sections of all reactions mean that
the difference in the evaporation residue cross sections for the $^{48}$Ca+$^{249-252}$Cf
reactions may appear in dependence of the survival probability of compound nucleus
and excited intermediate nuclei along the de-excitation  cascade of CN on
its mass number $A_{\rm CN}$.

\section{Evaporation residue cross section and discussion}
\label{sect3}

The excitation function of the individual evaporation residues (ER)
 along the de-excitation cascade of compound nuclei formed in the investigated
$^{48}$Ca+$^{249,250.251,252}$Cf reactions are calculated by formula
(see Refs. \cite{FazioMPL2005,Fazio05})
\begin{equation}
\label{evapor}
\sigma_{\rm ER}^{(x)}(E^*_x)=\sum_{\ell=0}^{\ell_d}(2\ell+1)
\sigma^{(x-1)}(E^*_x,\ell)W_{\rm sur}^{(x-1)}(E^*_x,\ell),
\end{equation}
where $\sigma^{(x-1)}(E^*_x,\ell)$ is the partial formation cross-section
of the excited intermediate nucleus of the $(x-1)$th step
and $W_{\rm sur}^{(x-1)}(E^*_x,\ell)$ is the survival probability of the
$(x-1)$th intermediate nucleus against fission along the de-excitation
cascade of CN. It is clear that $\sigma^{(0)}(E^*_x,\ell)=\sigma_{\rm fus}(E^*_x,\ell)$.
 The maximum value $\ell_{\rm d}$  of partial waves contributing to capture
events is found by solving the equations of motion (1)-(7) for the given initial values of
the energy $E_{\rm c.m.}$ and orbital angular momentum $\ell_0$ of  collision,
 at each value orientation angle $\alpha_T$ of the axial symmetry of the
deformed target nucleus (in details see Ref. \cite{NuclPhys05}).
We should stress that the real number of partial waves contributing
to the ER formation is much smaller than $\ell_{\rm d}$ because
$W_{\rm sur}^{(x-1)}(E^*_x,\ell)$ depends on the fission barrier
being a sum of the parameterized macroscopic fission
barrier $B_{fis}^{m}(\ell)$  depending on the angular momentum $J$
 and  the microscopic (shell) correction
$\delta W$
\begin{equation}
\label{fissb} B_{fis}(\ell,T)=c \ B_{fis}^{m}(\ell)-h(T) \ q(\ell) \ \delta
W.
\end{equation}
In our calculations, superheavy isotopes of element $Z=118$  have not macroscopic barrier,
$B_{fis}^{m}(\ell)=0$ and we took into account damping of the shell correction
by increasing the excitation energy $E^*_x$  and $\ell$ angular momentum
of fissioning nucleus by the functions $h(T)$ and $q(\ell)$, respectively.
These functions are
\begin{equation}
h(T) = \{ 1 + \exp [(T-T_{0})/ d]\}^{-1} 
\label{hoft}
\end{equation}
 and 
\begin{equation}
q(\ell) = \{ 1 + \exp [(\ell-\ell_{1/2})/\Delta \ell]\}^{-1},
\label{hofl}
\end{equation}

where, in Eq. (\ref{hoft}), $d= 0.3$~MeV is the rate of washing out the shell corrections with
the temperature, and $T_0=1.16$~MeV is the value at which the damping factor $h(T )$
is reduced by 1/2; analogously, in Eq. (\ref{hofl}), $\Delta \ell =3\hbar$ is the rate of washing out the shell corrections with the angular momentum, and $\ell_{1/2} =20\hbar$ is the value
at which the damping factor $q(\ell)$ is reduced by 1/2.
 This
procedure allows the shell corrections to become dynamical
quantities, also.
 Therefore, if the capture of $^{48}$Ca by an isotope of Cf takes place up to
values $\ell_{\rm d}=105$ the fission barrier disappears
at $\ell > 40$ due to damping the shell correction by $q(\ell)$.

The partial cross section of complete fusion is calculated by formula
(see Refs. \cite{FazioMPL2005,Fazio05} )
 \begin{eqnarray}
 \label{totfus}
 \sigma_{\rm fus}(E_{\rm c.m.},\ell;\beta_P, \alpha_T) & = &
\sigma_{\rm cap}(E_{\rm c.m.},\ell;\beta_P, \alpha_T)\times\nonumber\\ & \times &
 P_{CN}(E_{\rm c.m.},\ell; \beta_P, \alpha_T),
 \end{eqnarray}

\begin{figure}[h]
\vspace*{2.6cm}
\begin{center}
\resizebox{0.55\textwidth}{!}{\includegraphics{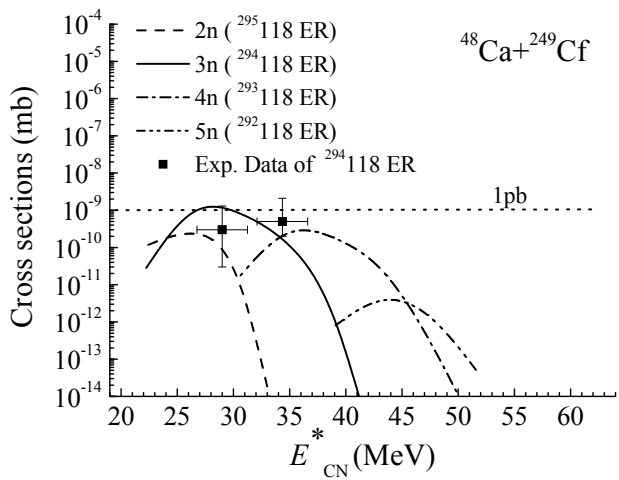}}
\vspace*{-3.00 cm} \caption{\label{fig8} Individual evaporation residue excitation
functions after emission of 2 (dashed line), 3 (full line), 4 (dash-dotted line),
and 5 (dash-double dotted line) neutrons  from the $^{297}$118 CN in the
$^{48}$Ca+$^{249}$Cf reaction, by using in calculation the masses and fission
barrier values of Refs. \cite{,Muntian03,Kowal}.
The  experimental data (full squares) of the $^{294}$118 ER formation cross section
obtained from Ref.\cite{Ogan06}.}
\end{center}
\end{figure}

In Fig. \ref{fig8} we report the ER cross sections for the $^{48}$Ca+$^{249}$Cf reaction
after emission of 2, 3, 4, and 5 neutrons from the $^{297}$118 CN as a
function of the $E^*_{\rm CN}$ excitation energy. The calculated
ER cross sections were obtained by using the mass and fission barrier values
given in Refs. \cite{Muntian03,Kowal} of the Warsaw group.

Since the fission barrier component of the macroscopic rotating liquid drop model
is zero for the formed superheavy nuclei, the component caused by the shell effects
(microscopic model) is damped by a function depending on the nuclear temperature
and angular momentum of CN ( see Ref. \cite{FazioJPSJ72}). In this figure
we  present the data obtained in the $^{48}$Ca+$^{249}$Cf experiment reported
in Ref.  \cite{Ogan06} regarding the synthesis of the $^{294}$118 superheavy
 nucleus obtained after 3 neutron emission from the $^{297}$118 CN, at two projectile
energies corresponding to excitation energies of $E^*=29.2$ and 34.4 MeV of the
compound nucleus. As Fig. \ref{fig8}, shows the maximum values of cross sections
connected with the 2n, 3n, and 4n  emission channels are included in the
$0.3-1.2$ pb range.  It is possible, in principle,  to detect the $^{295}$118,
$^{294}$118, and $^{293}$118 evaporation residue nuclei  which formed
after emission of  2, 3, and 4 neutrons,  respectively, from the  $^{297}$118 CN,
at convenient $^{48}$Ca beam energies in the $241-253$ MeV interval.
In this figure, the result of the $^{294}$118 evaporation excitation function (3n channel)
is  in fairly good agreement with the experimental  data of Ref. \cite{Ogan06}.
In fact, the calculated values of the ER cross sections at $E^*_{\rm CN}$=29.2 and 34.4 MeV
are close to the experimental data barely within the error bars.
For this reason we decided to continue analysis and interpret the origination
of difference between our results and  observed experimental data (see in forward
Fig. \ref{fig14}).

\begin{figure}[h]
\vspace*{2.6cm}
\begin{center}
\resizebox{0.55\textwidth}{!}{\includegraphics{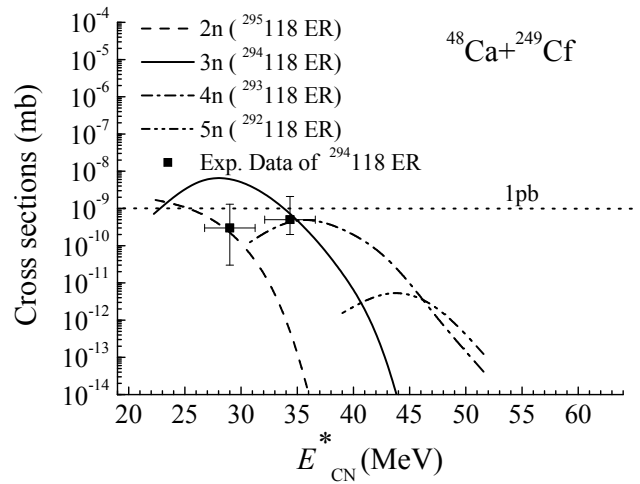}}
\vspace*{-3.00 cm} \caption{\label{fig9} As Fig. \ref{fig8}, but by using in  calculation
the masses  of Ref. \cite{MolNix} and  fission barriers of Ref. \cite{Mol09}.}
\end{center}
\end{figure}

\begin{figure}[h]
\vspace*{2.6cm}
\begin{center}
\resizebox{0.55\textwidth}{!}{\includegraphics{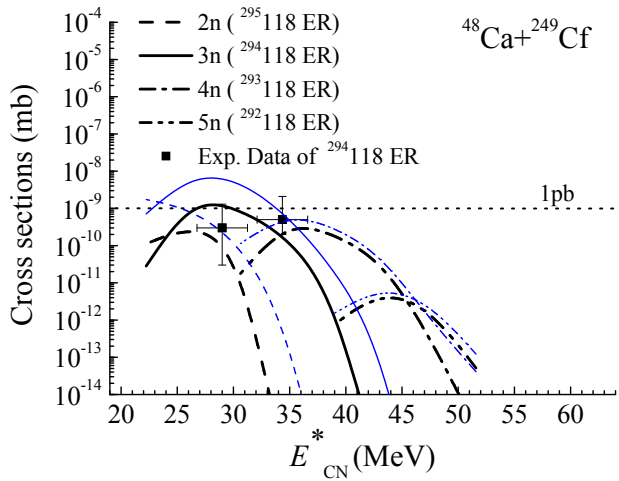}}
\vspace*{-3.00 cm} \caption{\label{fig10} (Color online) Comparison of results reported in
Figs. \ref{fig8} and \ref{fig9}. Thick lines are obtained by using the values of Refs. \cite{Muntian03,Kowal},
thin lines by using the values of Refs. \cite{MolNix,Mol09}.}
\end{center}
\end{figure}

In Fig. \ref{fig9} we report the analogous results as shown in Fig. \ref{fig8}, obtained
for the same $^{48}$Ca+$^{249}$Cf reaction leading to the $^{297}$118 CN, but by using in
 the calculation the masses of Ref.  \cite{MolNix} and  the fission barriers of Ref. \cite{Mol09} given by M\"oller {\it et. al}. As the figure shows, the excitation function of the 3n
evaporation channel is in good agreement with the second experimental point only,
but in general the excitation functions of evaporation residue nuclei  by this way
are higher than the  results obtained by using the masses \cite{Muntian03} and
fission barriers \cite{Kowal} of the Warsaw group. The comparison of  the results
calculated by the both set of theoretical masses and barriers with the experimental data
from Ref. \cite{Ogan06} is shown in Fig. \ref{fig10}. All  results obtained
by using the two different masses and barriers data were performed with the same set of other parameters above-described in this paper.

 In the following Figs. \ref{fig11}, \ref{fig12} and \ref{fig13}
we compared the ER excitation functions obtained in this work by using
the masses and barriers of Refs. \cite{Muntian03,Kowal} (thick lines)
and Refs. \cite{MolNix,Mol09}  (thin lines) for
the other $^{48}$Ca+$^{250,251,252}$Cf investigated reactions leading to the
$^{298}$118, $^{299}$118, and $^{300}$118 compound nuclei, respectively.
The maximum values of the ER excitation functions for the 3n emission
channel reach or overestimate  10 pb when the mass and fission barrier
values of Refs. \cite{MolNix,Mol09} are used, moreover,
these values are larger than experimental data of Ref. \cite{Ogan06}
 and more higher than the corresponding values which have been
found when the masses and barriers of Refs. \cite{Muntian03,Kowal}
are used. Therefore, in the following analysis we choose to refer
to the excitation functions (thick lines)
obtained by using masses and barriers of Refs. \cite{Muntian03,Kowal}
because the results appear more closer to the
experimental data.

\begin{figure}[h]
\vspace*{2.6cm}
\begin{center}
\resizebox{0.55\textwidth}{!}{\includegraphics{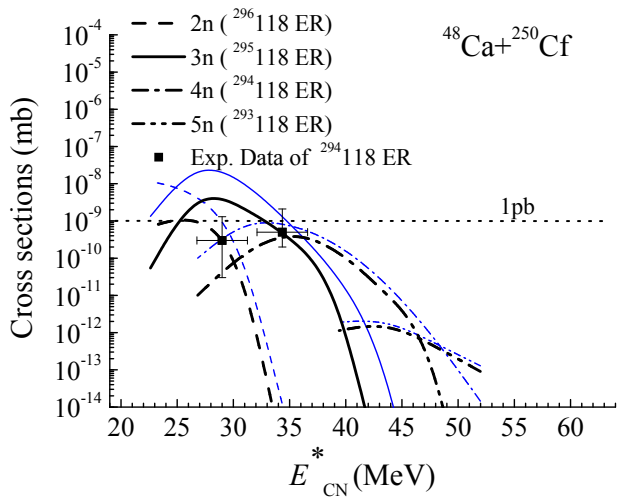}}
\vspace*{-3.00 cm} \caption{\label{fig11} (Color online) As Fig. \ref{fig10}, but for the
$^{48}$Ca+$^{250}$Cf reaction.}
\end{center}
\end{figure}

As one can see in Fig. \ref{fig11},  the cross section calculated for the
$^{294}$118 evaporation residue nucleus, which is obtained
after emission of 4 neutrons from the $^{298}$118 CN  being a
product of the $^{48}$Ca+$^{250}$Cf reaction, is also in
good agreement with the data of the $^{294}$118 evaporation residue
 synthesized in the $^{48}$Ca+$^{249}$Cf experiment after 3
neutron emission from $^{297}$118 CN.

 In the experimental identification of the ER nucleus by
the $\alpha$-decay chain assures
only the $^{294}$118 formation but the predecessor de-excitation
cascade--3 neutrons emission from the $^{297}$118 CN or 4 neutrons
emission from the $^{298}$118 CN can not be distinguished.
 The problem is that the presence of the $^{250}$Cf
isotope in the used target  in addition with the $^{249}$Cf isotope
 is inevitable and therefore it is necessary to take into account the $^{294}$118
contributions caused by the both  $^{48}$Ca+$^{249}$Cf and
$^{48}$Ca+$^{250}$Cf reactions.

\begin{figure}[h]
\vspace*{2.6cm}
\begin{center}
\resizebox{0.55\textwidth}{!}{\includegraphics{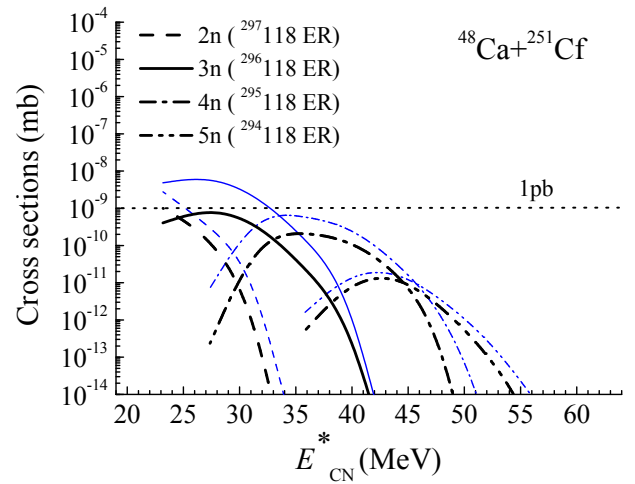}}
\vspace*{-3.00 cm} \caption{\label{fig12} (Color online) As Fig. \ref{fig10}, but for the
$^{48}$Ca+$^{251}$Cf reaction.}
\end{center}
\end{figure}

\begin{figure}[h]
\vspace*{1.6cm}
\begin{center}
\resizebox{0.55\textwidth}{!}{\includegraphics{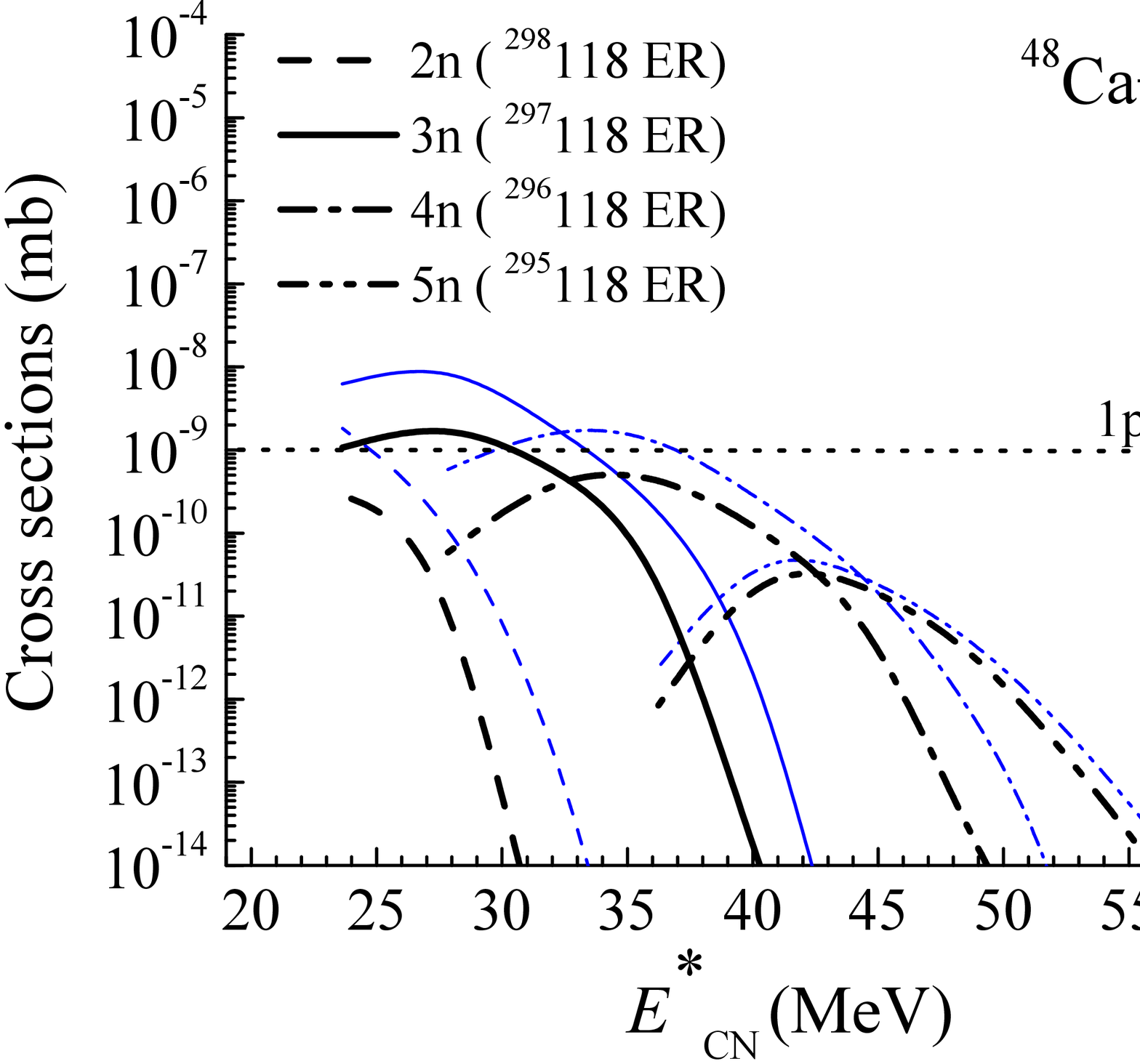}}
\vspace*{-2.60 cm} \caption{\label{fig13} (Color online) As Fig. \ref{fig10}, but for the
$^{48}$Ca+$^{252}$Cf reaction.}
\end{center}
\end{figure}

\begin{figure}[h]
\vspace*{2.6cm}
\begin{center}
\resizebox{0.55\textwidth}{!}{\includegraphics{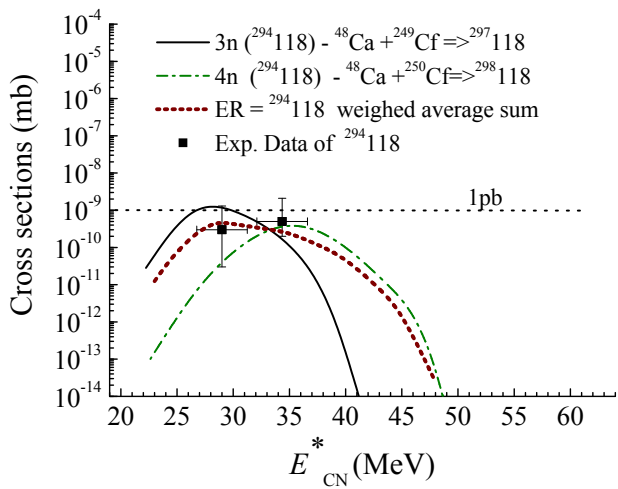}}
\vspace*{-3.00 cm} \caption{\label{fig14} (Color online) Contributions of the $^{294}$118
evaporation residue nucleus synthesized by the $^{48}$Ca+$^{249}$Cf (full line)
and $^{48}$Ca+$^{250}$Cf  (dash-dotted line) reactions. The dotted line represents the
weighed average sum of the two mentioned contributions.}
\end{center}
\end{figure}

In Fig. \ref{fig14} we report the excitation functions of the 3n
evaporation channel in the reaction with the $^{249}$Cf isotope and
4n evaporation channel in the reaction with $^{250}$Cf isotope, as
well as the experimental data from Ref. \cite{Ogan06}. In this
figure we add the weighed two excitation functions of the same
$^{294}$118 evaporation residue nucleus reached by both
$^{48}$Ca+$^{249,250}$Cf reactions after 3 and 4 neutron emission,
respectively (see Figs. \ref{fig8} and \ref{fig11}),  represented by
the dotted line in Fig. \ref{fig14}.   As one can see, this averaged
excitation function of the $^{294}$118 formation is in complete
agreement with the data of Ref. \cite{Ogan06}.

As regards the possibility to detect  also the formation of $^{295}$118 ER
 nucleus obtained after 3n emission from the  $^{298}$118 CN in the $^{48}$Ca+$^{250}$Cf
reaction (see Fig. \ref{fig11}), we observe that this ER nucleus reaches the maximum
yield of about
4 pb at  $E^*_{\rm CN}=28$~MeV  of the $^{298}$118 CN. Moreover,
at $E^*_{\rm CN}=35$~MeV the $^{295}$118 ER formation reaches the
appreciable value of 0.6 pb obtained as sum of the contribution due to
the 3 neutron emission from the $^{298}$118 CN in the $^{48}$Ca+$^{250}$Cf
reaction (see Fig. \ref{fig11}) and also the contribution due to the $^{295}$118
ER formation after 4 neutron emission from $^{299}$118 CN in the  $^{48}$Ca+$^{251}$Cf
reaction  (see Fig. \ref{fig12}).
Moreover, as one can see in Fig. \ref{fig13}  the population of the 3n-channel (corresponding
to the $^{297}$118 ER formation) should reach about 2 pb at the $E^*_{\rm CN}$ excitation
energy of 27-28 MeV of the  $^{300}$118 CN while the population of
 the 4n-channel (leading to the $^{296}$118 ER) is  near  0.5 pb at the $E^*_{\rm CN}$
excitation energy of about 35 MeV of the $^{300}$118 CN. Therefore,
the $^{48}$Ca+$^{252}$Cf reaction also appears as an useful and accessible reaction
in order to obtain the $^{297}$118 and $^{296}$118 ER nuclei after 3n and 4n emission from
the $^{300}$118 CN, respectively.

One can see in Figs. \ref{fig10}-\ref{fig13}  that  the highest yield of ER nuclei
formation in the $E^*_{\rm CN}=25-40$~MeV excitation energy interval is obtained for
the 3n-channel of the four investigated reaction, but the 2n- and 4n-channel
are  also populated by an appreciable mode.
Of course, the experimental data observed in the  $^{48}$Ca+$^{249}$Cf reaction
\cite{Ogan06}  corresponding to the formation of the $^{294}$118 ER at
$E^*_{\rm CN}=29.2$ and 34.4 MeV  may be meaningfully
considered as contributions of the 3n-channel of the $^{48}$Ca+$^{249}$Cf reaction and
 the 4n-channel of the $^{48}$Ca+$^{250}$Cf reaction.
If we assume a hypothesis that the target contains also the $^{251}$Cf and $^{252}$Cf
 isotopes, we can verify that the contributions of the $^{294}$118
ER nucleus formed in the $^{48}$Ca+$^{251,252}$Cf reactions are very small
(lower than 10$^{-9}$ and 10$^{-6}$ pb, respectively) in the above-mentioned
29.2 - 34.4 MeV excitation energy interval of the corresponding CN,
because the $^{294}$118 ER nucleus should be reached after 5n emission
from the $^{299}$118 CN, and 6n emission from the $^{300}$118 CN.
Therefore, in the excitation energy range
$29.2-34.4$~MeV, only the events of the 4n emission from the
$^{298}$118 CN formed in the $^{48}$Ca+$^{250}$Cf reaction can
contribute to the population of the $^{294}$118 ER nucleus (see thick dashed
line in Fig. \ref{fig11}) in addition to the contribution of the 3n
emission from the $^{297}$118 CN formed in the $^{48}$Ca+$^{249}$Cf
reaction (see full line in Fig. \ref{fig8}).

Moreover, as Figs. \ref{fig10},  \ref{fig11},  \ref{fig12}, and  \ref{fig13} show,
the evaporation
residue yields for the 2n, 3n, 4n, and 5n channels for the reactions with
even-even $^{250,252}$Cf targets are higher  than for the ones with
even-odd $^{249,251}$Cf targets.
 The ER excitation functions  calculated by using the masses and
barriers of Ref. \cite{MolNix} are higher than the ones obtained
by using the values  of Refs. \cite{Muntian03,Kowal}.
By the comparison of the excitation functions of ER in reactions leading
to the  $^{297-300}$118 CN's, we can affirm that the use of  masses and barriers
of Refs. \cite{Muntian03,Kowal} in calculation of  the evaporation residue nuclei
leads to the results which are close to the experimental data while the values obtained
from Ref. \cite{MolNix,Mol09} lead to overestimation them. Moreover, one can observe
from results from Figs. \ref{fig10}-\ref{fig13} that
the cross sections  determined for the ER nuclei obtained after
2n, 3n, and 4n emission from the $^{297-300}$118 CN's are included
 in the about $0.2-4$~pb range in the $E^*_{\rm CN}=25-40$~MeV excitation energy region
of the  corresponding CN which are formed in the $^{48}$Ca+$^{249-252}$Cf reactions.
Therefore, the events and related cross section  of the $^{293-298}$118 ER nuclei yields
in the  $^{48}$Ca+$^{249-252}$Cf reactions can be observed and measured.
The $^{293}$118 and $^{298}$118 ER nuclei can be only observed in the $^{48}$Ca+$^{249}$Cf and
$^{48}$Ca+$^{252}$Cf, respectively, while the  other  pairs $^{294,295}$118, $^{295,296}$118, and
$^{296,297}$118 ER nuclei can be observed in the   $^{48}$Ca+$^{249,250}$Cf,  $^{48}$Ca+$^{250,251}$Cf,
and $^{48}$Ca+$^{251,252}$Cf reactions, respectively, with respect to the presence of contiguous Cf isotopes
in target.
In this case, instead to work with a target enriched with one of Cf isotope only,
it is convenient to have one target constituted of two or more isotopes of Cf
(as for example $^{249,252}$Cf) because at various $^{48}$Ca beam energies
 the contiguous evaporation residue yields  produced by $x$n-channels of the
two  reactions can be explored.
The rate of ER contributions  depends on  the $^{48}$Ca beam energy
(and then with the $E^*_{\rm CN}$ energy) and also  on the peculiarities of the
ER formation channels. The presence of various contiguous isotopes in target gives
the possibility to observe different ER nuclei in the same experiment and
to compare the rate of various ER yields. In such a
case the preparation of the target is less expansive for cost and time,
and the set of  experimental data is more rich in yields and variety of registered ER nuclei.

In addition to our previous analysis and discussion,
it is useful to compare the results of the excitation functions
presented in Fig. \ref{fig12} of  this paper for the
$^{48}$Ca+$^{251}$Cf reaction  (when the values of
masses and fission barriers of  Refs. \cite{Muntian03,Kowal} are used) with the corresponding
results given in Fig. 3 (b) of Ref. \cite{Zag12} for  the same  reaction.
 Apart from the fact that the excitation functions of the main 3n- and 4n- channel of
Ref. \cite{Zag12} are about 2 and 3 times, respectively, higher than
 our corresponding values reported in Fig. \ref{fig12}.
 It  seems to be not realistic that the 2n- and
3n- excitation functions are peaked at 35 and 36 MeV of excitation energy, respectively.
 In the formation of evaporation residue with about 0.1 pb  cross section 2 emitted
neutrons take away 42 MeV from the $^{299}$118 CN against about 13 MeV requested for
the neutron binding energy of 2 neutrons.
 It means that each neutron should move with a
kinetic energy of about $14-15$ MeV while in average the neutron kinetic energy is close
to the  CN temperature in this case  which is about
0.9 MeV. Analogously for the 3n channel where at 0.1 pb of ER cross section the 3
emitted neutrons  take away 48 MeV from the compound nucleus against about 19.5  MeV
 requested for the neutron binding energy of 3 neutrons. Also in this case each
neutron moves with a kinetic energy of about 10 MeV in comparison with the nuclear
temperature of the compound nucleus that is about 1 MeV. That is an unrealistic result.
In fact, in our evaporation residue excitation functions reported in Figs.
\ref{fig8}-\ref{fig14}, the results always lead to an average neutron kinetic
energy of about 0.9 - 1 MeV.

 In conclusion of the present discussion,  we can affirm that our complete model is
able to describe
the evolution of  dinuclear system during reaction up to the
CN formation and CN's de-excitation cascade. This model leads to reliable
results of individual excitation functions of evaporation residue nuclei as a function
of energy and orbital angular momentum for each projectile-target combination (see Refs.
 \cite{jpcs2010,ijmf2010,jpcs2011}).

\section{Conclusions}
\label{sect4}

We investigated the formation of the  heaviest evaporation residue nuclei from
the $^{297-300}$118 CN which are formed in reactions induced by collision
of the $^{48}$Ca projectiles with the heaviest accessible actinide targets $^{249-252}$Cf.
If in future it will be possible to prepare targets of $^{254}$Es and $^{257}$Fm, then
the $^{302}$119  and $^{305}$120 CN's may be formed, but these targets are in every way  the extreme limit
of possibility of synthesizing SHE's by using the $^{48}$Ca beam
because other heavier Es and Fm nuclei as well as other heavier actinide nuclei are radioactive with shorter lifetimes. Therefore, it is impossible to prepare useful targets with the aim to synthesize   superheavy elements heavier than $^{302}$119 and $^{305}$120 by $^{48}$Ca induced reactions.

By analyzing the 2, 3, 4,  and 5  neutron emission channels along the de-excitation
cascade of compound nuclei formed in the  $^{48}$Ca+$^{249-252}$Cf reactions we studied
the possibilities of   synthesizing  the $^{292-298}$118  ER nuclei. In addition,
by considering the experimental conditions nowadays available in Laboratories, the more
convenient and accessible reaction channels of observing evaporation residue nuclei are
the 3 and 4 neutron emission channels in the $^{48}$Ca+$^{249-252}$Cf reactions at beam
energies corresponding to the $E^*=25-40$~MeV excitation energy range of compound nuclei.

Moreover, we found higher capture cross sections for the $^{48}$Ca+$^{250,252}$Cf reactions in comparison with the ones of the $^{48}$Ca+$^{249,251}$Cf reactions.
We discussed  the influence of the entrance channel dynamics on the capture, quasifission, and
fusion cross sections by considering the mass asymmetry parameter,
shell effects of reactants, dynamical deformation of nuclei in the DNS formation,
interaction angle between the axial symmetry axes at collision of projectile and
target nuclei. We also considered the effects of masses and fission barriers on the
evaporation residue nuclei when the values of Refs. \cite{Muntian03,Kowal} or the ones
of Ref. \cite{MolNix,Mol09} are used.

By comparing the results of our analysis regarding the study of the
$^{48}$Ca+$^{249,250}$Cf reactions with the  data obtained in the experiment of
Ref. \cite{Ogan06} regarding the observation of the $^{294}$118  evaporation residue
nucleus, we conclude that the better description of the experimental results is that
the observed $^{294}$118 synthesis events \cite{Ogan06} registered at two different
beam energies are contributed by the the 3n-channel in the $^{48}$Ca+$^{249}$Cf reaction
and 4n-channel in the $^{48}$Ca+$^{250}$Cf reaction, due to the inevitable presence of
the $^{250}$Cf isotope in the $^{249}$Cf enriched target.
Moreover, the comparison of results obtained for the ER nuclei in the investigation of the $^{48}$Ca+$^{252}$Cf reaction suggest to use one target only constituted of all the Cf isotopes of more long lifetimes. It is more convenient the procedure for its preparation, and in one experiment only it is possible to observe and  study a wide set of ER nuclei formed by 2n, 3n, 4n, and 5n emission channels, only changing the $^{48}$Ca beam energy in the about $E_{\rm lab}=235-260$~MeV range.

\begin{acknowledgments}
A.K. Nasirov is grateful to the Istituto Nazionale di Fisica Nucleare
 and Department of Physics of the University of Messina for
 the support received in the collaboration between the Dubna and
 Messina groups, and he thanks the Russian Foundation for Basic
 Research for the financial support in the performance of this
 work. A.~Sobiczewski would like to thank Yu.~Oganessian, J.~Roberto,and V.~Utyonkov 
for useful discussions about the questions connected with the possible use 
of the californium target , which should be a mixture of isotopes , in the
synthesis of the 118 element. A.~S. acknowledges a support by the Polish National Centre of Science (within the research project No. N N 202 204938) and the Polish-JINR (Dubna) Cooperation Programme.
\end{acknowledgments}

\end{document}